\documentclass[11pt,a4paper]{article}
\pdfoutput=1

\usepackage[utf8]{inputenc}
\usepackage[T1]{fontenc}
\usepackage[margin=2.5cm]{geometry}
\usepackage[pdfstartview={FitH},linktoc=page,colorlinks=true,linkcolor=blue,citecolor=blue,urlcolor=blue]{hyperref}
\usepackage[numbered]{bookmark}
\usepackage{mathtools}
\usepackage{amssymb}
\usepackage{textcomp}
\usepackage{graphicx}
\usepackage[font=small,labelfont=bf]{caption}
\usepackage{authblk}
\usepackage{cite}
\usepackage{dsfont}
\usepackage{xcolor}
\usepackage[titles]{tocloft}
\usepackage{setspace}

\numberwithin{equation}{section}
\allowdisplaybreaks

\newcommand*{\dd}{\mathop{}\!\mathrm{d}}

\DeclarePairedDelimiter\bra{\langle}{\rvert}
\DeclarePairedDelimiter\ket{\lvert}{\rangle}
\DeclarePairedDelimiterX\braket[2]{\langle}{\rangle}{#1 \delimsize\vert #2}

\DeclareMathOperator{\csch}{csch}

\begin{document}


\title{\vspace{1.5cm}\textbf{Entanglement of higher-derivative oscillators in holographic systems}\vspace{1cm}}
\author[a]{Hristo Dimov}
\author[a]{Stefan Mladenov}
\author[a,b]{Radoslav C. Rashkov}
\author[a]{Tsvetan Vetsov}
\affil[a]{\textit{Department of Physics, Sofia University,}\authorcr
	\textit{5 J. Bourchier Blvd., 1164 Sofia, Bulgaria}\vspace{5pt}}
\affil[b]{\textit{Institute for Theoretical Physics, Vienna University of Technology,}\authorcr
	\textit{Wiedner Hauptstr. 8--10, 1040 Vienna, Austria}\vspace{10pt}}
\affil[ ]{\texttt{\href{mailto:h_dimov@phys.uni-sofia.bg}{h\_dimov},\href{mailto:smladenov@phys.uni-sofia.bg}{smladenov},\href{mailto:rash@phys.uni-sofia.bg}{rash},\href{mailto:vetsov@phys.uni-sofia.bg}{vetsov@phys.uni-sofia.bg}}\vspace{1cm}}
\date{}
\maketitle

\begin{abstract}
We study the quantum entanglement of coupled Pais-Uhlenbeck oscillators using the formalism of thermo-field dynamics. The entanglement entropy is computed for the specific cases of two and a ring of $N$ coupled Pais-Uhlenbeck oscillators of fourth order. It is shown that the entanglement entropy depends on the temperatures, frequencies and coupling parameters of the different degrees of freedom corresponding to harmonic oscillators. We also make remarks on the appearance of instabilities of higher-derivative oscillators in the context of AdS/CFT correspondence. Finally, we advert to the information geometry theory by calculating the Fisher information metric for the considered system of coupled oscillators.
\end{abstract}

\vspace{1.5cm}

\thispagestyle{empty}
\newpage


\noindent\rule{\linewidth}{0.75pt}
\vspace{-0.8cm}\tableofcontents\vspace{0.1cm}
\noindent\rule{\linewidth}{0.75pt}


\section{Introduction}

Fundamental laws of physics are usually governed by at most second order differential equations, the reason for that being quite clear. Ostrogradsky's approach \cite{Ostrogradsky1850} to higher-derivative theories produces terms in the Hamiltonian, which are linear in particles' momenta. This imminently leads to an unbounded from below Hamiltonian at classical level, hence instabilities and ghost problem after quantisation. One should not, however, rashly label a higher-derivative theory unphysical having in mind the following argument. Consider a system of two second-order differential equations which gives stable dynamics. Generally, this system is equivalent to one differential equation of fourth order. It may happen that the Ostrogradsky's approach gives boundless negative energy spectrum, which does not necessarily mean unstable system. The explanation is that the Ostrogradsky's energy may not coincide with the energy given by the system of two second-order differential equations.

The Pais-Uhlenbeck oscillator (PUO) \cite{Pais:1950za} has been proposed as a prototype of higher-derivative theories of gravity, which are renormalisable \cite{PhysRevD.16.953}. This simple mechanical model has been subject to renewed interest in recent years due to many works inferring that the PUO can be regarded as a viable physical system. Owing to the presence of higher-order time derivatives, the PUO cannot be quantised using the canonical approach. Instead one uses the Dirac constraints method to construct the correct Hamiltonian, which is diagonal with positive and negative states \cite{Mannheim:2004qz}. Taking a limit corresponding to equal frequencies, in which the second-order derivative vanishes leaving pure fourth-order PUO, the negative norm states go off-shell thus purging the theory.

The first step towards quantisation of the PUO is a suitable Hamiltonian definition, which improves the Ostrogradsky's one. The search for proper Hamiltonian is based on the observation that there always exists at least two quadratic in variables integrals of motion \cite{2005AcPPB..36.2115B}. Since the Hamiltonian is also an integral of motion, one can make a general ansatz for it. Consequent requirement of equivalence between the corresponding equations of motion and the initial Lagrange-Euler equations leads to a derivation of Poisson-like structure. Thus the efforts to find the right Hamiltonian resulted in a bunch of alternative Hamiltonian formulations of the PUO \cite{Mostafazadeh:2010yw,Masterov:2015ija,Masterov:2016jft}. Furthermore, the stability of this system has been studied using both analytical and numerical methods \cite{2013MPLA...2850038P,Pavsic:2013noa,Pavsic:2013mja}.

Another interesting problem is the question of entangled states. Ever since 1935, entanglement \cite{PhysRev.47.777,Schrodinger1935} has been challenging our understanding of quantum phenomena. Taking advantage of the presence of well-defined Hamiltonian formulation of the PUO giving viable physics, in this paper we are interested in studying the quantum entanglement of systems of such oscillators. One convenient way to deal with this problem is to use the framework of thermo-field dynamics (TFD) of a double Hilbert space \cite{2013PhyA..392.3518H}. In this context, it has been shown recently that the entanglement entropy of coupled harmonic oscillators depends on the temperature and coupling parameters \cite{Nakagawa:2016ady}. Alternatively, the entanglement entropy can be derived from holography \cite{Ryu:2006bv,Ryu:2006ef} (for a review see \cite{Nishioka:2009un}). The idea is analogous to the Bekenstein-Hawking entropy of black holes. Specifically, the entanglement entropy of $(d+1)$-dimensional conformal field theory is related by the AdS/CFT correspondence to the area of a static $d$-dimensional minimal surface in $\text{AdS}_{d+2}$. Hence, one can study certain aspects of string theory on microscopic level by making use of thermodynamic quantities such as quantum entanglement entropy.

The rest of this paper is structured as follows. In section \ref{sec:oscHolo} we focus on the appearance of higher-derivative oscillators in the context of the AdS/CFT correspondence. We emphasise the importance of the knowledge one can gain within this approach for understanding of particular corners of information space in this context. In section \ref{sec:two1D4PUOs} we consider two minimally coupled PUOs and compute their entanglement entropy. In section \ref{sec:N1D4PUOs} we refer to the more general case of a ring of PUOs and compute the entanglement entropy of one of the oscillators (first subsystem) and the others (second subsystem). We finalise our discussion with some concluding remarks in section \ref{sec:conclusion}. In appendix \ref{sec:BCM} we set aside some technical details concerning the Hamiltonian diagonalisation.

\section{Appearance of higher-derivative oscillators in holographic models}
\label{sec:oscHolo}

\subsection{Reduction of gauge theories with CS terms to PUOs}
\label{subsec:oscCS}

Higher-derivative theories naturally arise in string theory in the context of holography. One beautiful example of a holographic correspondence is the equivalence between 3D topological Chern-Simons gauge theory and the chiral half of a rational conformal field theory \cite{Witten1:1989, Axelrod:1991}. It appears as a special case of the celebrated AdS/CFT correspondence, namely the case of superstring theories on space-times of the form ${\rm{AdS}}_3\times K^7$, where $K^7$ is some compact space. It is shown that the low-energy supergravity theory on ${\rm{AdS}}_3$ typically contains gauge fields with Chern-Simons terms. The relevance of topological field theories to the gauge/gravity correspondence was first discussed in \cite{Witten2:1998}. Besides in string theory, the importance of massive Chern-Simons theories have also been recognized in condensed matter physics with a view towards quantum computations \cite{Freedman:2004}.

In this subsection we would like to show that under some assumptions the massive Chern-Simons theories naturally reduce to higher-derivative oscillators, namely Pais-Uhlenbeck oscillators. To demonstrate this point we will elaborate on the simplest example of a topologically massive electrodynamics described by the Lagrangian \cite{Schonfeld:1981, Deser:1982, Deser2:1999}:
\begin{equation}\label{topmassive ED}
{\cal L} =  - \frac{1}{4}\,{F_{\mu \nu }}\,{F^{\mu \nu }} - {\varepsilon ^{\mu \nu \lambda }}\,\left( {\frac{m}{4}\,{A_\mu }\,{F_{\nu \lambda }} + \frac{k}{m}\,{F_{\mu \sigma }}\,{\partial ^\sigma }{F_{\nu \lambda }}} \right)\,,\quad {F_{\mu \nu }} = {\partial _\mu }{A_\nu } - {\partial _\nu }{A_\mu }\,,
\end{equation}
where $k$ is a dimensionless constant. The  given (2+1)-dimensional system (\ref{topmassive ED}) can be reduced to a mechanical one via the substitution $A^\mu(x)\to\sqrt{m}\,r^\mu(t)$ and $B^2=m^2\,e^{-1}$. The resulting  Lagrangian is
\begin{equation}\label{mechlag}
L = \frac{1}{2}\,m\,\dot x_i^2 + \frac{{e\,B}}{2}\,{\varepsilon _{ij}}\,{x_i}\,{{\dot x}_j} + k\,{\varepsilon _{ij}}\,{{\dot x}_i}\,{{\ddot x}_j}\,.
\end{equation}
In (\ref{topmassive ED}) we impose the Weyl gauge, $A^0=0$, which  effectively leads to the elimination of the variable $x^0$. Thus it is evident that the Lagrangian (\ref{mechlag}) describes a charged higher-derivative oscillator coupled to an external homogeneous magnetic field $B$. The equations of motion of the effective theory are given by
\begin{align}
&x_1^{(3)}(t) - \frac{{B\,e}}{{2\,k}}\,{{\dot x}_1}(t) - \frac{m}{{2\,k}}\,{{\ddot x}_2}(t) = 0\,,\\
&x_2^{(3)}(t) - \frac{{B\,e}}{{2\,k}}\,{{\dot x}_2}(t) + \frac{m}{{2\,k}}\,{{\ddot x}_1}(t) = 0\,.
\end{align}
Making the substitution $y_i(t)=\dot x_i(t)$ this system transforms to a fourth-order PU oscillator:
\begin{equation}\label{4thorderPUOeq}
y_i^{(4)}(t) + \,\left( {\omega _1^2 + \omega _2^2} \right)\,{{\ddot y}_i}(t) + \omega _1^2\,\omega _2^2\,{y_i}(t) = 0\,.
\end{equation}
The oscillator frequencies are related to the parameters of the mechanical system by the following expressions:
\begin{equation}
\omega _1^2 + \omega _2^2 = \frac{{{m^2}}}{{4\,{k^2}}} - \frac{{B\,e}}{k},\,\quad \omega _1^2\,\omega _2^2 = \frac{{{B^2}\,{e^2}}}{{4\,{k^2}}}\,.
\end{equation}
The system (\ref{mechlag}), however, as  most higher-derivative models, has a spectrum unbounded from below. This can be cured by supplying the coupled system with an appropriate constraints \cite{Horvathy:2005}. Classically this is equivalent to replacing the higher-derivative Lagrangian
(\ref{mechlag}) with the first-order exotic Duval-Horvathy Lagrangian \cite{Duval:2000,Duval:2005},
\begin{equation}\label{HD exotica}
L = {P_i}\,\dot x_i^{} - \frac{1}{{2\,m}}\,P_i^2 + \frac{\theta }{2}\,{\varepsilon _{ij}}\,{P_i}\,{{\dot P}_j} + \frac{{e\,B}}{2}\,{\varepsilon _{ij}}\,{x_i}\,{{\dot x}_j}\,,
\end{equation}
which generates equations of motion with the effective mass $\mu  = m\,(1 - e\,B\,\theta )$, ${P_i} = \mu \,{{\dot x}_i}$, ${{\dot P}_i} = e\,B\,{\varepsilon _{ij}}\,{{\dot x}_j}$. The exotic particle system (\ref{HD exotica}) can also be obtained by a reduction of another (2+1)-dimensional Abelian gauge field theory given by the Lagrangian with several Chern-Simons terms \cite{Hagen:1987, Deser3:1987},
\begin{equation}
{\cal L} = {\varepsilon ^{\mu \nu \lambda }}\,{\Phi _\mu }\,{\partial _\nu }{A_\lambda } - \frac{\lambda }{2}{\Phi _\mu }\,{\Phi ^\mu } - \frac{k}{{2\,m}}\,{\varepsilon ^{\mu \nu \lambda }}\,{\Phi _\mu }\,{\partial _\nu }{\Phi _\lambda } - \frac{{\beta \,m}}{2}\,{\varepsilon ^{\mu \nu \lambda }}\,{A_\mu }\,{\partial _\nu }{A_\lambda }\,.
\end{equation}

In the next subsection we turn our attention to the appearance of higher-derivative oscillators in another interesting holographic model, namely the Pilch-Warner supergravity solution.

\subsection{Higher-derivative oscillators and the Pilch-Warner solution}
\label{subsec:oscPWsol}

The Pilch-Warner (PW) flow geometry \cite{Pilch:2000ej,Pilch:2000fu} is a solution of five-dimensional $\mathcal{N}=8$ supergravity lifted to ten dimensions. It interpolates between two critical points of the scalar potential. In the ultraviolet (UV) critical point the solution reproduces the maximally supersymmetric $\text{AdS}_5\times S^5$ type IIB background. In the infrared (IR) critical point it gives another type IIB supergravity solution, which represents a warped $\text{AdS}_5$ times a squashed five-sphere geometry. The solution preserves $1/8$ of the supersymmetry all over the flow, except for the IR point where the supersymmetry is enhanced to $1/4$. In this section we will focus on the IR point, in which the gravitational theory corresponds by holography to a large $N$ limit of the $\mathcal{N}=1$ superconformal theory of Leigh-Strassler \cite{Leigh:1995ep}.

We will now show that higher-derivative oscillators naturally appear in the Penrose limit of the Pilch-Warner solution in the presence of $B$-field. The pp-wave limit of the PW solution in its IR point is investigated in \cite{Brecher:2002ar} for a particular choice of null geodesic, which corresponds to the moduli space of a D3-brane probe. The relevant fields that will contribute the bosonic part of the world-sheet action are the metric and the NS-NS antisymmetric two-form field $B_2$. For a gauge choice, in which the $B$-field permeates in the $X^1$, $X^2$, $X^3$, $X^4$, and $U$ directions, the action takes the form
\begin{align}
	S_B=-\frac{1}{4\pi\alpha'}\int\dd\tau\dd\sigma&\left[\sqrt{-g}g^{\alpha\beta}
	\left(2\partial_\alpha U\partial_\beta V+A_{ij}X^iX^j\partial_\alpha U\partial_\beta U
	+\partial_\alpha X^i\partial_\beta X^i\right)\right.\nonumber\\
	&\left.-2\sqrt{3}E\epsilon^{\alpha\beta}\left(X^1\partial_\alpha U\partial_\beta X^3
	-X^2\partial_\alpha U\partial_\beta X^4\right)\right],
\end{align}
where $\epsilon^{01}=1$ and the pp-wave spectrum matrix is $A_{ij}=\text{diag}[1,1,4]$. The quantity $E$ is the conserved energy corresponding to the Killing vector $\partial/\partial\tau$. A convenient world-sheet gauge is $g_{\alpha\beta}=\eta_{\alpha\beta}$ and the world-sheet coordinates are $\sigma^0=\tau$, $\sigma^1=\sigma$. The equation of motion for $U$ is $\nabla^2U=0$. In light-cone gauge $U=\alpha'p^+\tau+\text{const}$, the equations of motion for the world-sheet scalars are:
\begin{align}
\label{eq:EoMsys}
	\nabla^2X^1-4M^2X^1+\sqrt{3}M\partial_\sigma X^3&=0,\nonumber\\
	\nabla^2X^2-4M^2X^2-\sqrt{3}M\partial_\sigma X^4&=0,\nonumber\\
	\nabla^2X^3-M^2X^3-\sqrt{3}M\partial_\sigma X^1&=0,\nonumber\\
	\nabla^2X^4-M^2X^4+\sqrt{3}M\partial_\sigma X^2&=0,\nonumber\\
	\nabla^2X^p-M^2X^p&=0,
\end{align}
where $p=5,6,7,8$ are the directions unaffected by the $B$-field and $M=E\alpha'p^+$. By making use the ansatz $X^i(\tau,\sigma)=e^{i\sigma}x_i(\tau),~i=1,\dotsc,8$, the system of second-order partial differential equations \eqref{eq:EoMsys} boils down to the following equivalent system:
\begin{align}
\label{eq:EoMPUO}
	x_q^{(4)}+\left(5M^2+2\right)x_q^{(2)}+\left(4M^4+2M^2+1\right)x_q&=0,\nonumber\\
	x_p^{(2)}+\left(M^2+1\right)x_p&=0,
\end{align}
where $q=1,2$ and the derivative is taken with respect to $\tau$. Let us stress that the crucial point here is the inclusion of the Kalb-Ramond two-form $B_2$. Its role is evident from the form of the system of partial differential equations \eqref{eq:EoMsys}---the $B$-field ties up the equations for $X^1,~X^3$ and $X^2,~X^4$ thus making the system equivalent to one, in which some of the differential equations are of higher order. Therefore we conclude that the presence of $B$-field results in dynamics, which can be modeled by higher-derivative Pais-Uhlenbeck oscillators.

On the other hand one can consider quadratic fluctuations around classical solutions. The case of rotating strings in PW background is considered in \cite{Dimov:2003bh}. Since we expect that the Kalb-Ramond field will play again significant role, we will consider only the $S^5$ part of the geometry where the $B$-field is turned on. The Lagrangian, describing the quadratic fluctuations, is
\begin{align}
	\mathcal{L}_{S^5}^\text{tot}=\partial_\alpha\tilde{\zeta}^A\partial^\alpha\tilde{\zeta}^A
	+\tilde{M}^2\left(\tilde{\zeta}_{\tilde{2}}^2+\tilde{\zeta}_{\tilde{4}}^2\right)
	&+\left(4\tilde{M}^2+\frac{3}{2}\bar{\rho'}^2\right)
	\left(\tilde{\zeta}_{\tilde{3}}^2+\tilde{\zeta}_{\tilde{5}}^2\right)\nonumber\\
	&+2\sqrt{3}\tilde{M}\left(\tilde{\zeta}_{\tilde{4}}\partial_1\tilde{\zeta}_{\tilde{5}}
	-\tilde{\zeta}_{\tilde{2}}\partial_1\tilde{\zeta}_{\tilde{3}}\right),
\end{align}
where $\tilde{M}^2=\frac{4}{9}(c_\beta+c_\gamma+c_\phi)^2$ is a constant depending on the angular velocities $c_\beta$, $c_\gamma$, and $c_\phi$ of the string rotating in the directions $\beta$, $\gamma$, and $\phi$, correspondingly\footnote{For further details, please see \cite{Dimov:2003bh}.}. The corresponding equations of motion are:
\begin{align}
	\nabla^2\tilde{\zeta}_{\tilde{1}}&=0,\nonumber\\
	\nabla^2\tilde{\zeta}_{\tilde{2}}-\tilde{M}^2\tilde{\zeta}_{\tilde{2}}+\sqrt{3}\tilde{M}
	\partial_1\tilde{\zeta}_{\tilde{3}}&=0,\nonumber\\
	\nabla^2\tilde{\zeta}_{\tilde{3}}-\left(4\tilde{M}^2+\frac{3}{2}\bar{\rho'}^2\right)
	\tilde{\zeta}_{\tilde{3}}-\sqrt{3}\tilde{M}\partial_1\tilde{\zeta}_{\tilde{2}}&=0,\nonumber\\
	\nabla^2\tilde{\zeta}_{\tilde{4}}-\tilde{M}^2\tilde{\zeta}_{\tilde{4}}-\sqrt{3}\tilde{M}
	\partial_1\tilde{\zeta}_{\tilde{5}}&=0,\nonumber\\
	\nabla^2\tilde{\zeta}_{\tilde{5}}-\left(4\tilde{M}^2+\frac{3}{2}\bar{\rho'}^2\right)
	\tilde{\zeta}_{\tilde{5}}+\sqrt{3}\tilde{M}\partial_1\tilde{\zeta}_{\tilde{4}}&=0,
\end{align}
where $\bar{\rho'}^2$ is $\sigma$-dependent. For the case of point-like string the above equations are equivalent to the pp-wave limit equations \eqref{eq:EoMsys}. Using the ansatz $\tilde{\zeta}_{\tilde{i}}=e^{i\sigma}y_i(\tau)$ leads to the following system of differential equations:
\begin{align}
\label{eq:EoMPUO2}
	y_p^{(4)}+\left[5\tilde{M}^2+2+\frac{3}{2}\bar{\rho}'^2\right]y_p^{(2)}
	+\left[4\tilde{M}^4+2\tilde{M}^2+1+\left(\tilde{M}^2+1\right)
	\frac{3}{2}\bar{\rho'}^2\right]y_p&=0,\nonumber\\
	y_1^{(2)}+y_1&=0,
\end{align}
where $p=2,4$. The first equation in \eqref{eq:EoMPUO2} is slightly different than the EoM of fourth-order PUO since $\bar{\rho'}^2$ is a function of $\sigma$. However, one can consider two limiting cases, i.e. short and long strings. In short-string approximation $\bar{\rho'}^2\approx1/\!\left(2\coth^2\rho_0-2\right)$, while in long-string approximation $\bar{\rho'}^2\approx\frac{1}{\pi^2}\log^2\left(\coth^2\rho_0-1\right)$, $\rho_0$ being the maximal value of $\rho(\sigma)$ \cite{Dimov:2003bh}. Hence, the coefficient in front of $y_p$ is constant in these two approximations and we again arrive at the fourth-order Pais-Uhlenbeck oscillator.
\\
\indent In order to elaborate on the consequences of these results in the following sections we will proceed with a detailed analysis of the fourth-order PUOs given in Eq. (\ref{4thorderPUOeq}).

\section{Two minimally coupled 1D fourth-order PU oscillators}
\label{sec:two1D4PUOs}

As a first step towards description of higher-derivative oscillators in the context of information geometry, we will consider two minimally coupled one-dimensional fourth-order Pais-Uhlenbeck oscillators. Following the Ostrogradsky's approach, the dynamics of such system is governed by an alternative Hamiltonian of decoupled harmonic oscillators \cite{Masterov:2015ija}\footnote{For an alternative Hamiltonian formulation of the odd-order PUO, please see \cite{Masterov:2016jft}.} supplemented with interaction term:
\begin{equation}
\label{eq:H2}
	H_2=\frac{1}{2}\sum_{\mu=1}^2\sum_{k=0}^1\mathrm{sgn}(\alpha_{\mu,k})
	\left(p_\mu^kp_\mu^k+\omega_{\mu,k}^2x_\mu^kx_\mu^k\right)
	+\frac{1}{2}\sum_{\mu\neq\nu=1}^2c_{\mu\nu}x_\mu x_\nu,
\end{equation}
where the index $\mu$ labels the different oscillators of fourth-order, any of them with distinct frequencies $\omega_{\mu,k}$. Since the oscillators are one-dimensional, we do not include spatial index in the canonical variables $\left(x_\mu^k,p_\mu^k\right)$. The canonical coordinates $\left(x_\mu^k,p_\mu^k\right)$ corresponding to the Poisson structure are defined through \cite{Pais:1950za}
\begin{align}
\label{eq:cancoord}
	x_\mu^0&=\sqrt{\frac{|\alpha_{\mu,0}|}{\omega_{\mu,1}^2-\omega_{\mu,0}^2}}
	\left(x_\mu^{(2)}+\omega_{\mu,1}^2x_\mu\right),&
	p_\mu^0&=\mathrm{sgn}(\alpha_{\mu,0})\frac{\dd x_\mu^0}{\dd t},\nonumber\\
	x_\mu^1&=\sqrt{\frac{|\alpha_{\mu,1}|}{\omega_{\mu,1}^2-\omega_{\mu,0}^2}}
	\left(x_\mu^{(2)}+\omega_{\mu,0}^2x_\mu\right),&
	p_\mu^1&=\mathrm{sgn}(\alpha_{\mu,1})\frac{\dd x_\mu^1}{\dd t}.
\end{align}
Consequently, the coordinate $x_\mu$ is expressed in terms of the canonical coordinates as
\begin{equation}
	x_\mu=\frac{\sqrt{|\alpha_{\mu,1}|}x_\mu^0-\sqrt{|\alpha_{\mu,0}|}x_\mu^1}
	{\sqrt{|\alpha_{\mu,1}||\alpha_{\mu,0}|\left(\omega_{\mu,1}^2-\omega_{\mu,0}^2\right)}}.
\end{equation}
Let us consider the physically interesting case, in which all constants $\alpha_{\mu,k}$ are positive, i.e. $\mathrm{sgn}(\alpha_{\mu,k})=1$. This choice will keep us away from the ghost problem after quantisation. Thus the Hamiltonian \eqref{eq:H2} can be represented as a sum of free and interaction parts, $H_2=H_2^\mathrm{F}+H_2^\mathrm{I}$, each of them given by:
\begin{equation}
	H_2^\mathrm{F}=\frac{1}{2}\sum_{\mu=1}^2\sum_{k=0}^1p_\mu^kp_\mu^k,\quad
	H_2^\mathrm{I}=\frac{1}{2}\sum_{\mu=1}^2\sum_{k=0}^1\omega_{\mu,k}^2x_\mu^kx_\mu^k
	+\sum_{k,l=0}^1\tilde{c}^{kl}x_1^kx_2^l,
\end{equation}
where the new coupling constant is
\begin{equation}
	\tilde{c}^{kl}=(-1)^{k+l}c_{12}\left[\alpha_{1,k}\,\alpha_{2,l}
	\left(\omega_{1,1}^2-\omega_{1,0}^2\right)\left(\omega_{2,1}^2-\omega_{2,0}^2\right)\right]^{-1/2}.
\end{equation}
The next step is to simultaneously diagonalise the kinetic and potential parts of the Hamiltonian. This will be done by introducing uniform coordinates in the phase space $\eta=\left(x_\mu^k,p_\mu^k\right)^\mathrm{T}$ and writing the Hamiltonian in matrix form
\begin{equation}
	H_2=\frac{1}{2}\eta^\mathrm{T}
	\begin{pmatrix}
		\Omega & 0\\
		0 & \mathds{1}_4
	\end{pmatrix}
	\eta,
\end{equation}
where the matrix $\Omega$ is given by
\begin{equation}
	\Omega=
	\begin{pmatrix}
		\omega_{1,0} & 0 & \tilde{c}^{00} & \tilde{c}^{01}\\
		0 & \omega_{1,1} & \tilde{c}^{10} & \tilde{c}^{11}\\
		\tilde{c}^{00} & \tilde{c}^{10} & \omega_{2,0} & 0\\
		\tilde{c}^{01} & \tilde{c}^{11} & 0 & \omega_{2,1}
	\end{pmatrix}.
\end{equation}
In the case of identical PUOs the frequencies are equal in pairs, i.e. $\omega_{1,0}=\omega_{2,0}=\omega_0$ and $\omega_{1,1}=\omega_{2,1}=\omega_1$, and the coefficients $\tilde{c}^{kl}$ become symmetric in their two indices. In order to simplify the notations we rename the coefficients as follows: $\tilde{c}^{00}=c_0$, $\tilde{c}^{11}=c_1$, $\tilde{c}^{01}=\tilde{c}^{10}=c_2$. Thus the matrix $\Omega$ turns into symmetric block circulant matrix with symmetric blocks:
\begin{equation}
\label{eq:OmegaWC}
	\Omega=
	\begin{pmatrix}
		W & C\\
		C & W
	\end{pmatrix},
	\quad W=
	\begin{pmatrix}
		\omega_0 & 0\\
		0 & \omega_1\\
	\end{pmatrix},
	\quad C=
	\begin{pmatrix}
		c_0 & c_2\\
		c_2 & c_1
	\end{pmatrix}.
\end{equation}
Now we can apply the procedure described in appendix \ref{sec:BCM}. The number of blocks that define the symmetric block circulant matrix $\Omega$ are $n=2$, hence $j=0,1$ and we obtain $\mathbf{H_0}=W+C$, $\mathbf{H_1}=W-C$. The corresponding eigenvalues $\lambda_{0\pm}$, $\lambda_{1\pm}$ and eigenvectors $\mathbf{v}_0$, $\mathbf{v}_1$ are given by the expressions:
\begin{align}
\label{eq:freq}
	\lambda_{0\pm}&=a_+\pm c_2h_+\rightarrow\mathbf{v}_{0\pm}=(b_+\pm h_+,1)^\mathrm{T},\nonumber\\
	\lambda_{1\pm}&=a_-\pm c_2h_-\rightarrow\mathbf{v}_{1\pm}=(b_-\pm h_-,1)^\mathrm{T},
\end{align}
where
\begin{align}
	a_\pm&=\frac{1}{2}\left[\omega_0+\omega_1\pm(c_0+c_1)\right],\nonumber\\
	b_\pm&=\frac{1}{2c_2}\left[c_0-c_1\pm(\omega_0-\omega_1)\right],\nonumber\\
	h_\pm&=\frac{1}{2c_2}\sqrt{2c_2^2+\left[c_0-c_1\pm(\omega_0-\omega_1)\right]^2}.
\end{align}
One should keep in mind that the parameters $\omega_0$, $\omega_1$, $c_0$, and $c_1$ are not independent because the eigenvalues $\lambda_{0\pm}$ and $\lambda_{1\pm}$ must be real.

The matrix $P$, which diagonalises the matrix $\Omega$, is constructed from the normalised $\mathbf{w}_j/||\mathbf{w}_j||$ eigenvectors \eqref{eq:eigenvecw} and takes the form
\begin{equation}
	P=\frac{1}{\sqrt{2}}
	\begin{pmatrix}
		\frac{b_++h_+}{\sqrt{1+(b_++h_+)^2}} & \frac{b_+-h_+}{\sqrt{1+(b_+-h_+)^2}} &
		\frac{b_-+h_-}{\sqrt{1+(b_-+h_-)^2}} & \frac{b_--h_-}{\sqrt{1+(b_--h_-)^2}}\\
		\frac{1}{\sqrt{1+(b_++h_+)^2}} & \frac{1}{\sqrt{1+(b_+-h_+)^2}} &
		\frac{1}{\sqrt{1+(b_-+h_-)^2}} & \frac{1}{\sqrt{1+(b_--h_-)^2}}\\
		\frac{b_++h_+}{\sqrt{1+(b_++h_+)^2}} & \frac{b_+-h_+}{\sqrt{1+(b_+-h_+)^2}} &
		\frac{-b_--h_-}{\sqrt{1+(b_-+h_-)^2}} & \frac{-b_-+h_-}{\sqrt{1+(b_--h_-)^2}}\\
		\frac{1}{\sqrt{1+(b_++h_+)^2}} & \frac{1}{\sqrt{1+(b_+-h_+)^2}} &
		\frac{-1}{\sqrt{1+(b_-+h_-)^2}} & \frac{-1}{\sqrt{1+(b_--h_-)^2}}
	\end{pmatrix}.
\end{equation}
Finally, taking into account that the matrix $P$ is orthogonal ($P^{-1}=P^\mathrm{T}$), we obtain the diagonalised Hamiltonian
\begin{equation}
\label{eq:H2diag}
	H_2=\frac{1}{2}\eta^\mathrm{T}
	\begin{pmatrix}
		PDP^\mathrm{T} & 0\\
		0 & \mathds{1}_4
	\end{pmatrix}
	\eta=\frac{1}{2}\widehat{\eta}^\mathrm{T}
	\begin{pmatrix}
		D & 0\\
		0 & \mathds{1}_4
	\end{pmatrix}
	\widehat{\eta},
\end{equation}
where the coordinate transformations and the diagonal matrix $D$ are given by:
\begin{equation}
	\widehat{\eta}=
	\begin{pmatrix}
		P^\mathrm{T} & 0\\
		0 & \mathds{1}_4
	\end{pmatrix}
	\eta,
	\quad D=\text{diag}[\lambda_{0+},\lambda_{0-},\lambda_{1+},\lambda_{1-}].
\end{equation}
Furthermore, we define creation and annihilation operators $\mathfrak{a}_j^\dagger$ and $\mathfrak{a}_j$, $j=1,\ldots,4$, as
\begin{align}
	\mathfrak{a}_j&=\frac{1}{\sqrt{2\hbar}}\left(\sqrt{\lambda_j}\widehat{x}_j
	+\frac{i}{\sqrt{\lambda_j}}\widehat{p}_j\right),\nonumber\\
	\mathfrak{a}_j^\dagger&=\frac{1}{\sqrt{2\hbar}}\left(\sqrt{\lambda_j}\widehat{x}_j
	-\frac{i}{\sqrt{\lambda_j}}\widehat{p}_j\right),
\end{align}
where $\widehat{x}_j$ and $\widehat{p}_j$ are the coordinate and momentum part of the transformed vector $\widehat{\eta}$, and the frequencies $\lambda_j$ are connected to the diagonal elements of the matrix $D$ by $[\lambda_1,\lambda_2,\lambda_3,\lambda_4]=[\lambda_{0+},\lambda_{0-},\lambda_{1+},\lambda_{1-}]^{1/2}$. The creation and annihilation operators satisfy the commutation relations
\begin{equation}
	[\mathfrak{a}_j,\mathfrak{a}_k]=[\mathfrak{a}_j^\dagger,\mathfrak{a}_k^\dagger]=0,\quad
	[\mathfrak{a}_j,\mathfrak{a}_k^\dagger]=\delta_{jk}.
\end{equation}
Thus the Hamiltonian \eqref{eq:H2diag} takes the standard form
\begin{equation}
\label{eq:H2quant}
	H_2=\sum_{j=1}^4\hbar\lambda_j\left(\mathfrak{a}_j^\dagger\mathfrak{a}_j+\frac{1}{2}\right).
\end{equation}
The Fock space is built up from the vacuum in the usual way,
\begin{equation}
	\ket*{\{n_j\}}=\prod_{j=1}^4\frac{(\mathfrak{a}_j^\dagger)^{n_j}}{\sqrt{n_j!}}\ket*{\{0\}},
\end{equation}
where $\ket*{\{n_j\}}=\ket*{n_1}\otimes\cdots\otimes\ket*{n_4}$. The energy spectrum is derived from the eigenvalue problem $H_2\ket*{\{n_j\}}=\mathcal{E}_{\{n_j\}}\ket*{\{n_j\}}$,
\begin{equation}
	\mathcal{E}_{\{n_j\}}=\sum_{j=1}^4\hbar\lambda_j\left(n_j+\frac{1}{2}\right).
\end{equation}
In terms of the excited states, which are orthonormal $\braket{\{m_j\}}{\{n_j\}}=\delta_{\{m_j\},\{n_j\}}$, the diagonal Hamiltonian \eqref{eq:H2quant} can be written in matrix form
\begin{equation}
	H_2=\sum_{\{n_j\}=0}^\infty\sum_{j=1}^4\hbar\lambda_j\left(n_j+\frac{1}{2}\right)\ket*{\{n_j\}}\bra*{\{n_j\}}.
\end{equation}
The partition function of the system is defined as
\begin{equation}
	Z(K_j):=\mathrm{Tr}_{\{j\}}e^{-\beta H_2}=
	\prod_{j=1}^4\frac{\exp(-K_j/2)}{1-\exp(-K_j)},
\end{equation}
where $\beta$ is the inverse temperature and $K_j=\hbar\lambda_j\beta$. Then the standard density matrix in equilibrium, $\rho_\text{eq}(K_j)=e^{-\beta H_2}/Z(K_j)$, takes the form
\begin{equation}
	\rho_\text{eq}(K_j)=\frac{1}{Z(K_j)}\sum_{\{n_j\}=0}^\infty\exp\left[-\sum_{j=1}^4K_j
	\left(n_j+\frac{1}{2}\right)\right]\ket*{\{n_j\}}\bra*{\{n_j\}}.
\end{equation}
Thermo-field dynamics explores a double Hilbert space defined as a direct product of two isomorphic Hilbert spaces---the original one with set of base vectors $\{\ket*{n}\}$ and its copy called the tilde space with bases $\{\ket*{\tilde{n}}\}$---namely $\{\ket*{n}\otimes\ket*{\tilde{m}}\}\equiv\{\ket*{n}\ket*{\tilde{m}}\}\equiv\{\ket*{n,\tilde{m}}\}$. Therefore one can define a statistical state by $\ket*{\Psi}=\sum_n\sqrt{\rho_\text{eq}}\ket*{n}\ket*{\tilde{n}}$, which can be easily computed for our system,
\begin{equation}
	\ket*{\Psi(K_j)}=\frac{1}{\sqrt{Z(K_j)}}\sum_{\{n_j\}=0}^\infty\exp\left[-\sum_{j=1}^4\frac{K_j}{2}
	\left(n_j+\frac{1}{2}\right)\right]\ket*{\{n_j\}}\ket*{\{\tilde{n}_j\}}.
\end{equation}
Using defined this way statistical state one can define extended density matrix by $\widehat{\rho}(K_j)=\ket*{\Psi(K_j)}\bra*{\Psi(K_j)}$:
\begin{equation}
	\widehat{\rho}(K_j)=\frac{1}{Z(K_j)}\sum_{\{n_j\}=0}^\infty\sum_{\{m_j\}=0}^\infty
	\exp\left[-\sum_{j=1}^4\frac{K_j}{2}\left(n_j+m_j+1\right)\right]
	\ket*{\{m_j\}}\bra*{\{n_j\}}\ket*{\{\tilde{m}_j\}}\bra*{\{\tilde{n}_j\}}
\end{equation}
Consequently, the renormalised extended density matrix $\widehat{\rho}_{1,2}(K_j)=\mathrm{Tr}_{3,4}\widehat{\rho}(K_j)$ of the first PUO (equivalently harmonic oscillators 1 and 2) is obtained by tracing out the second PUO (equivalently harmonic oscillators 3 and 4),
\begin{align}
	\widehat{\rho}_{1,2}(K_j)&=\frac{1}{Z(K_j)}
	\prod_{j=3}^4\frac{\exp(-K_j/2)}{1-\exp(-K_j)}\sum_{\substack{n_1,m_1\\n_2,m_2}}^\infty
	\exp\left[-\frac{K_1}{2}(n_1+m_1+1)-\frac{K_2}{2}(n_2+m_2+1)\right]\nonumber\\
	&\times\ket*{m_1,m_2}\bra*{n_1,n_2}\ket*{\tilde{m}_1,\tilde{m}_2}\bra*{\tilde{n}_1,\tilde{n}_2}.
\end{align}
Finally, one derives the extended entanglement entropy of the first PUO from the expression $\widehat{S}_{1,2}=-k_\text{B}\mathrm{Tr}_{1,2}\left[\widehat{\rho}_{1,2}\log\widehat{\rho}_{1,2}\right]$,
\begin{align}
\label{eq:entropy2}
	\widehat{S}_{1,2}(K_1,K_2)=\frac{k_\text{B}}{2}\coth\frac{K_1}{4}\coth\frac{K_2}{4}
	&\left[K_1\left(1+\coth\frac{K_1}{4}\right)+K_2\left(1+\coth\frac{K_2}{4}\right)\right.\nonumber\\
	&\left.-2\log\left[\left(e^{K_1}-1\right)\left(e^{K_2}-1\right)\right]\vphantom{\coth\frac{K_1}{4}}\right].
\end{align}
\begin{figure}[h]
	\centering
	\includegraphics[width=0.75\textwidth]{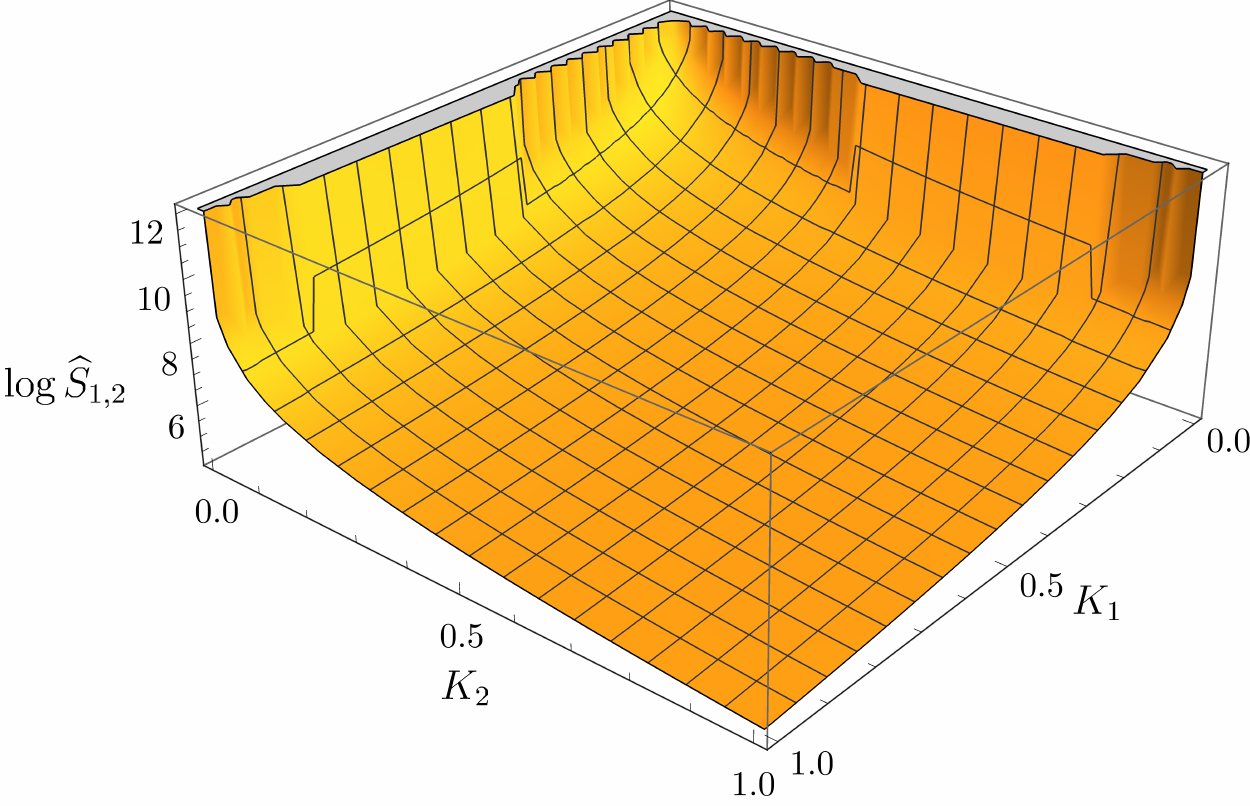}
	\caption{The entanglement entropy as function of $K_1$ and $K_2$ in units $k_\text{B}=1$.}
	\label{fig:entropy}
\end{figure}
The extended entanglement entropy $\widehat{S}_{1,2}$ as function of the inverse scaled temperatures $K_1$ and $K_2$ is visualised in figure \ref{fig:entropy} in units $k_\text{B}=1$. As can be seen from the figure, the entanglement entropy goes to infinity for large temperature (equivalent to small values of $K_1$ and $K_2$). Conversely, when the temperature approaches zero (corresponding to approaching infinity arguments of the function $\widehat{S}_{1,2}$), the entanglement entropy approaches zero, which is exactly the statement of Nernst's theorem.

\section{$N$ minimally coupled 1D fourth-order PU oscillators}
\label{sec:N1D4PUOs}

In this section we will slightly depart from the case considered in section \ref{sec:two1D4PUOs} by generalising the Hamiltonian \eqref{eq:H2} for description of $N$ coupled 1D fourth-order PUOs. Moreover, we will restrict ourselves to interaction only between the nearest neighbours over the chain. The generalisation to such Hamiltonian is obvious, namely
\begin{equation}
\label{eq:HN}
	H_N=\frac{1}{2}\sum_{\mu=1}^N\sum_{k=0}^1\mathrm{sgn}(\alpha_{\mu,k})
	\left(p_\mu^kp_\mu^k+\omega_{\mu,k}^2x_\mu^kx_\mu^k\right)
	+\frac{1}{2}\sum_{\langle\mu,\nu\rangle=1}^Nc_{\mu\nu}x_\mu x_\nu.
\end{equation}
Considering the case of physical significance, i.e. $\alpha_{\mu,k}>0$, one can split again the Hamiltonian to kinetic and interaction parts, $H_N=H_N^\mathrm{F}+H_N^\mathrm{I}$,
\begin{equation}
	H_N^\mathrm{F}=\frac{1}{2}\sum_{\mu=1}^N\sum_{k=0}^1p_\mu^kp_\mu^k,\quad
	H_N^\mathrm{I}=\frac{1}{2}\sum_{\mu=1}^N\sum_{k=0}^1\omega_{\mu,k}^2x_\mu^kx_\mu^k
	+\frac{1}{2}\sum_{\langle\mu,\nu\rangle=1}^N\sum_{k,l=0}^1\tilde{c}_{\mu\nu}^{kl}x_\mu^kx_\nu^l,
\end{equation}
where the constant $\tilde{c}_{\mu\nu}^{kl}$ is defined as
\begin{equation}
	\tilde{c}_{\mu\nu}^{kl}=(-1)^{k+l}c_{\mu\nu}\left[\alpha_{\mu,k}\,\alpha_{\nu,k}
	\left(\omega_{\mu,1}^2-\omega_{\mu,0}^2\right)\left(\omega_{\nu,1}^2-\omega_{\nu,0}^2\right)\right]^{-1/2}.
\end{equation}
Thus defined, the constants $\tilde{c}_{\mu\nu}^{kl}$ are symmetric under exchange of pair upper and lower indices, i.e. $\tilde{c}_{\mu\nu}^{kl}=\tilde{c}_{\nu\mu}^{lk}$. This symmetry has been lost in the definition of the corresponding constant in section \ref{sec:two1D4PUOs} because we explicitly did the sum over the two lower indices. We will focus our attention on a chain of identical PUOs, therefore we have to close the chain by imposing $\tilde{c}_{N\,N+1}^{kl}=\tilde{c}_{N1}^{kl}$. Introducing again the vector $\eta=(x_\mu^k,p_\mu^k)^\mathrm{T}$, the Hamiltonian can be written in matrix form as
\begin{equation}
	H_N=\frac{1}{2}\eta^\mathrm{T}
	\begin{pmatrix}
		\Omega & 0\\
		0 & \mathds{1}_{2N}
	\end{pmatrix}
	\eta,
\end{equation}
where the matrix $\Omega$ has a block tridiagonal structure
\begin{equation}
	\Omega=
	\begin{pmatrix}
		W_1 & C_1 & 0 & \cdots & C_N^\mathrm{T}\\
		C_1^\mathrm{T} & W_2 & C_2 & \cdots & 0\\
		0 & C_2^\mathrm{T} & \ddots & \ddots & \vdots\\
		\vdots & \vdots & \ddots & \ddots & C_{N-1}\\
		C_N & 0 & \cdots & C_{N-1}^\mathrm{T} & W_N
	\end{pmatrix}
\end{equation}
and $W_j$ and $C_j$ are square $2\times2$ matrices given by:
\begin{align}
	W_j&=
	\begin{pmatrix}
		\omega_{j,0} & 0\\
		0 & \omega_{j,1}
	\end{pmatrix},
	~j=1,\ldots,N;\\
	C_j&=
	\begin{pmatrix}
		\tilde{c}_{j\,j+1}^{0\,0} & \tilde{c}_{j\,j+1}^{0\,1}\\
		\tilde{c}_{j\,j+1}^{1\,0} & \tilde{c}_{j\,j+1}^{1\,1}
	\end{pmatrix},
	~j=1,\ldots,N.
\end{align}
Let us consider again the case of identical PUOs, namely $\omega_{1,k}=\omega_{2,k}=\ldots=\omega_{N,k}$ and $\alpha_{1,k}=\alpha_{2,k}=\ldots=\alpha_{N,k}$. As a consequence the coefficients $\tilde{c}_{\mu\nu}^{kl}$ become symmetric under exchange of any two upper or down indices, $\tilde{c}_{\mu\nu}^{kl}=\tilde{c}_{\mu\nu}^{lk}$. Hence the matrices $W_i$ are all equal to each other and the matrices $C_j=C_j^\mathrm{T}$ become symmetric and equal as well. Therefore the problem boils down to diagonalisation of the following $2N\times2N$ symmetric block circulant matrix with symmetric blocks
\begin{equation}
	\Omega=
	\begin{pmatrix}
		W & C & 0 & \cdots & C\\
		C & W & C & \cdots & 0\\
		0 & C & \ddots & \ddots & \vdots\\
		\vdots & \vdots & \ddots & \ddots & C\\
		C & 0 & \cdots & C & W
	\end{pmatrix},
\end{equation}
where $W$ and $C$ are defined as in \eqref{eq:OmegaWC}, and $\Omega\in\mathcal{BC}_{N,2}$. These type of matrices are diagonalised in terms of Discrete Fourier Transform (DFT). As a first step we will block diagonalise the matrix $\Omega$ by making use of an unitary matrix $U$,
\begin{equation}
	\widehat{\Omega}=U^{-1}\Omega U=\text{diag}[D_1,D_2,\ldots,D_N].
\end{equation}
Defining the matrix $U$ as build of the following diagonal $2\times2$ blocks
\begin{equation}
	U_{kl}=\frac{1}{\sqrt{N}}e^{2\pi ikl/N}\mathds{1}_2,\quad k,l=0,1,\ldots,N-1,
\end{equation}
we can treat the matrix $\Omega$ as standard circulant matrix, which is block diagonalised by the matrix $U$, and the diagonal elements of the block-diagonal matrix $\widehat{\Omega}$ consist of the eigenvalues of $\Omega$, namely
\begin{align}
	D_{k+1}&=W+C\rho_k+C\rho_k^{N-1}=W+\left(e^{2\pi ik/N}+e^{-2\pi ik/N}\right)C\nonumber\\
	&=W+2\cos\left(2\pi k/N\right)C.
\end{align}
Substituting with the expressions for $W$ and $C$ we obtain
\begin{equation}
	D_{k+1}=
	\begin{pmatrix}
		\omega_0+2\cos(k\theta)c_0 & 2\cos(k\theta)c_2\\
		2\cos(k\theta)c_2 & \omega_1+2\cos(k\theta)c_1
	\end{pmatrix},
\end{equation}
where $\theta=2\pi/N$ and $k=0,1,\ldots,N-1$. The only thing we need to do is to find the eigenvalues and eigenvectors of the matrices $D_{k+1}$. A straightforward calculation gives
\begin{equation}
	\lambda_{k+1\pm}=a_k\pm2c_2h_k\rightarrow\mathbf{v}_{k+1\pm}=(b_k\pm h_k,1)^\mathrm{T},
\end{equation}
where
\begin{align}
	a_k&=\frac{1}{2}\left[\omega_0+\omega_1+2{c_0+c_1}\cos(k\theta)\right],\nonumber\\
	b_k&=\frac{1}{4c_2}\left[2(c_0-c_1)+(\omega_0-\omega_1)\sec(k\theta)\right],\nonumber\\
	h_k&=\frac{1}{4c_2}\left[2(c_0-c_1)^2+8c_2^2+(\omega_0-\omega_1)^2
	+4(c_0-c_1)(\omega_0-\omega_1)\cos(k\theta)\right.\nonumber\\
	&\left.\hphantom{{}={}\frac{1}{4c_2}}+2\left((c_0-c_1)^2+4c_2^2\right)
	\cos(2k\theta)\right]^{1/2}\sec(k\theta).
\end{align}
The Hamiltonian \eqref{eq:HN} thus becomes (taking into account that the matrix $U$ is unitary, i.e. $U^{-1}=U^\dagger$, and $\eta^\mathrm{T}=\eta^\dagger$)
\begin{equation}
\label{eq:HNdiag}
	H_N=\frac{1}{2}\eta^\dagger\,\text{diag}\left[U\text{diag}[D_1,\ldots,D_N]
	U^\dagger,\mathds{1}_{2N}\right]\eta=\frac{1}{2}\widehat{\eta}^\dagger\,\text{diag}
	\left[\widehat{D}_1,\ldots,\widehat{D}_N,\mathds{1}_{2N}\right]\widehat{\eta},
\end{equation}
where the matrices $D_{k+1}$ are diagonalised by orthogonal matrices $P_{k+1}$ composed of the orthonormal vectors $\mathbf{v}_{k+1\pm}/||\mathbf{v}_{k+1\pm}||$, namely $D_{k+1}=P\widehat{D}_{k+1}P^{-1}$. The explicit form of the matrix $P_{k+1}$ is
\begin{equation}
	P_{k+1}=
	\begin{pmatrix}
		\frac{b_k+h_k}{\sqrt{1+(b_k+h_k)^2}} & \frac{b_k-h_k}{\sqrt{1+(b_k-h_k)^2}}\\
		\frac{1}{\sqrt{1+(b_k+h_k)^2}} & \frac{1}{\sqrt{1+(b_k-h_k)^2}}
	\end{pmatrix}.
\end{equation}
Finally, the diagonal matrices $\widehat{D}_{k+1}$ and the transformation between $\eta$ and $\widehat{\eta}$ are given by the following expressions:
\begin{align}
	\widehat{\eta}&=\text{diag}\left[\text{diag}[P_1^\dagger,\ldots,P_N^\dagger]
	U^\dagger,\mathds{1}_{2N}\right]\eta,\nonumber\\
	\widehat{D}_{k+1}&=\text{diag}[\lambda_{k+1+},\lambda_{k+1-}],\quad k=0,\ldots,N-1.
\end{align}
Furthermore, we define creation and annihilation operators $\mathfrak{a}_j^\dagger$ and $\mathfrak{a}_j$, $j=1,\ldots,2N$, as
\begin{align}
	\mathfrak{a}_j&=\frac{1}{\sqrt{2\hbar}}\left(\sqrt{\lambda_j}\widehat{x}_j
	+\frac{i}{\sqrt{\lambda_j}}\widehat{p}_j\right),\nonumber\\
	\mathfrak{a}_j^\dagger&=\frac{1}{\sqrt{2\hbar}}\left(\sqrt{\lambda_j}\widehat{x}_j
	-\frac{i}{\sqrt{\lambda_j}}\widehat{p}_j\right),
\end{align}
where $\widehat{x}_j$ and $\widehat{p}_j$ are the coordinate and momentum part of the transformed vector $\widehat{\eta}$, and the frequencies $\lambda_j$ are connected to the diagonal elements of the matrix $\widehat{D}_{k+1}$ by $\lambda_{2j-1}=\sqrt{\lambda_{j-1+}}$ and $\lambda_{2j}=\sqrt{\lambda_{j-1-}}$. The creation and annihilation operators satisfy the commutation relations
\begin{equation}
	[\mathfrak{a}_j,\mathfrak{a}_k]=[\mathfrak{a}_j^\dagger,\mathfrak{a}_k^\dagger]=0,\quad
	[\mathfrak{a}_j,\mathfrak{a}_k^\dagger]=\delta_{jk}.
\end{equation}
Thus the Hamiltonian \eqref{eq:HNdiag} takes the standard form
\begin{equation}
\label{eq:HNquant}
	H_N=\sum_{j=1}^{2N}\hbar\lambda_j\left(\mathfrak{a}_j^\dagger\mathfrak{a}_j+\frac{1}{2}\right).
\end{equation}
The Fock space is built up from the vacuum in the usual way,
\begin{equation}
	\ket*{\{n_j\}}=\prod_{j=1}^{2N}\frac{(\mathfrak{a}_j^\dagger)^{n_j}}{\sqrt{n_j!}}\ket*{\{0\}},
\end{equation}
where $\ket*{\{n_j\}}=\ket*{n_1}\otimes\cdots\otimes\ket*{n_{2N}}$. The energy spectrum is derived from the eigenvalue problem $H_N\ket*{\{n_j\}}=\mathcal{E}_{\{n_j\}}\ket*{\{n_j\}}$,
\begin{equation}
	\mathcal{E}_{\{n_j\}}=\sum_{j=1}^{2N}\hbar\lambda_j\left(n_j+\frac{1}{2}\right).
\end{equation}
In terms of the excited states, which are orthonormal $\braket{\{m_j\}}{\{n_j\}}=\delta_{\{m_j\},\{n_j\}}$, the diagonal Hamiltonian \eqref{eq:HNquant} can be written in matrix form
\begin{equation}
	H_N=\sum_{\{n_j\}=0}^\infty\sum_{j=1}^{2N}\hbar\lambda_j\left(n_j+\frac{1}{2}\right)\ket*{\{n_j\}}\bra*{\{n_j\}}.
\end{equation}
The partition function of the system is defined as
\begin{equation}
	Z(K_j):=\mathrm{Tr}_{\{j\}}e^{-\beta H_N}=
	\prod_{j=1}^{2N}\frac{\exp(-K_j/2)}{1-\exp(-K_j)},
\end{equation}
where $\beta$ is the inverse temperature and $K_j=\hbar\lambda_j\beta$. Following the same line of calculations as in section \ref{sec:two1D4PUOs}, we first compute the ordinary density matrix,
\begin{equation}
	\rho_\text{eq}=\frac{1}{Z(K_j)}\sum_{\{n_j\}=0}^\infty\exp\left[-\sum_{j=1}^{2N}K_j
	\left(n_j+\frac{1}{2}\right)\right]\ket*{\{n_j\}}\bra*{\{n_j\}}.
\end{equation}
Taking into account the different range of the sum, the statistical state is
\begin{equation}
	\ket*{\Psi(K_j)}=\frac{1}{\sqrt{Z(K_j)}}\sum_{\{n_j\}=0}^\infty\exp\left[-\sum_{j=1}^{2N}\frac{K_j}{2}
	\left(n_j+\frac{1}{2}\right)\right]\ket*{\{n_j\}}\ket*{\{\tilde{n}_j\}},
\end{equation}
and the obtained from it extended density matrix reads
\begin{equation}
	\widehat{\rho}(K_j)=\frac{1}{Z(K_j)}\sum_{\{n_j\}=0}^\infty\sum_{\{m_j\}=0}^\infty
	\exp\left[-\sum_{j=1}^{2N}\frac{K_j}{2}\left(n_j+m_j+1\right)\right]
	\ket*{\{m_j\}}\bra*{\{n_j\}}\ket*{\{\tilde{m}_j\}}\bra*{\{\tilde{n}_j\}}.
\end{equation}
The next step is to factorise the whole system into two subsystems. Since all PUOs over the ring are absolutely equivalent, we pick up one of them (for definiteness, we choose the PUO described by harmonic oscillators 1 and 2) to be the first subsystem thus all the rest being the second subsystem. Therefore the renormalised extended density matrix with respect to the first subsystem is calculated by taking trace over the second subsystem of harmonic oscillators $3,4,\ldots,2N$:
\begin{align}
	\widehat{\rho}_{1,2}(K_j)&=\frac{1}{Z(K_j)}
	\prod_{j=3}^{2N}\frac{\exp(-K_j/2)}{1-\exp(-K_j)}\sum_{\substack{n_1,m_1\\n_2,m_2}}^\infty
	\exp\left[-\frac{K_1}{2}(n_1+m_1+1)-\frac{K_2}{2}(n_2+m_2+1)\right]\nonumber\\
	&\times\ket*{m_1,m_2}\bra*{n_1,n_2}\ket*{\tilde{m}_1,\tilde{m}_2}\bra*{\tilde{n}_1,\tilde{n}_2}.
\end{align}
Finally, we end up with the extended entanglement entropy of a Pais-Uhlenbeck oscillator inside a ring of $N-1$ other oscillators,
\begin{align}
\label{eq:entropyN}
	\widehat{S}_{1,2}(K_1,K_2)=\frac{k_\text{B}}{2}\coth\frac{K_1}{4}\coth\frac{K_2}{4}
	&\left[K_1\left(1+\coth\frac{K_1}{4}\right)+K_2\left(1+\coth\frac{K_2}{4}\right)\right.\nonumber\\
	&\left.-2\log\left[\left(e^{K_1}-1\right)\left(e^{K_2}-1\right)\right]\vphantom{\coth\frac{K_1}{4}}\right].
\end{align}
At first look expressions \eqref{eq:entropy2} and \eqref{eq:entropyN} seem to be the same since they completely match. However, this is not the case because the frequencies in the definitions of the inverse scaled temperatures $K_1$ and $K_2$ are given by different expressions for the formulae \eqref{eq:entropy2} and \eqref{eq:entropyN}. Of course the behaviour of the entanglement entropy \eqref{eq:entropyN} as function of the inverse scaled temperature mimics that in figure \ref{fig:entropy} and the Nernst theorem is applicable again.

\section{Concluding remarks}
\label{sec:conclusion}

We have considered the quantum entanglement of coupled fourth-order Pais-Uhlenbeck oscillators in the context of thermo-field dynamics. Being defined by Lagrangian which contains higher-than-first time derivatives in variables, such theory is characterised by Ostrogradsky instabilities after passing to Hamiltonian formalism. However, there exist many alternative Hamiltonian formulations of the PUO, which prove that it can be considered as a reliable physical system. One of these formulations states that every PUO of order $k$ is equivalent to a system of $k$ harmonic oscillators with alternating signs or even more general---with the signs of arbitrary non-zero constants coming from the ansatz of the Hamiltonian as a sum of Noether currents. By choosing all constants to be positive, we already have positive-definite Hamiltonian without negative-norm states. It turns out that such Hamiltonian is very convenient for studying quantum entanglement between PUOs.

Possessing well-defined Hamiltonian system, we further introduced simple linear interaction between the closest-neighbour PUOs. This led to the necessity of diagonalisation of symmetric block circulant matrices with symmetric blocks, the matrix form of the Hamiltonian interaction part being such type of matrix. With diagonalised Hamiltonian in hands, we were able to apply the TFD framework after canonical quantisation and building the Fock space of the quantum system. The double Hilbert space has proven to be extremely useful in defining and exploring various statistical quantities as extended density matrix and extended statistical state, because the isomorphic copy of the original Hilbert space somehow played the role of a tracer. Consequently, we computed the extended entanglement entropy of one PUO with the rest of the system under consideration and showed that the obtained physical quantity obeys the Nernst heat theorem.

Let us now comment on the consequences of the results concerning the holographic models. In section \ref{sec:two1D4PUOs} we obtained the diagonalised Hamiltonian \eqref{eq:H2diag} with the new frequencies \eqref{eq:freq} written in terms of the parameters of the two fourth-order PUOs:
\begin{align}
	\lambda_{0\pm}&=\frac{1}{2}\left[\omega_0+\omega_1+(c_0+c_1)\right]\pm
	\frac{1}{2}\sqrt{2c_2^2+\left[c_0-c_1+(\omega_0-\omega_1)\right]^2},\nonumber\\
	\lambda_{1\pm}&=\frac{1}{2}\left[\omega_0+\omega_1-(c_0+c_1)\right]\pm
	\frac{1}{2}\sqrt{2c_2^2+\left[c_0-c_1-(\omega_0-\omega_1)\right]^2}.
\end{align}
Written this way, the Hamiltonian \eqref{eq:H2diag} represents an alternative Hamiltonian formulation of two free Pais-Uhlenbeck oscillators each of fourth order with frequencies $\lambda_{0\pm}$ and $\lambda_{1\pm}$ and corresponding equations of motion\footnote{For more information, please see \cite{Masterov:2015ija}.}:
\begin{align}
\label{eq:PUOeqs}
	x_1^{(4)}+\left(\lambda_{0+}^2+\lambda_{0-}^2\right)x_1^{(2)}
	+\lambda_{0+}^2\lambda_{0-}^2x_1&=0,\nonumber\\
	x_2^{(4)}+\left(\lambda_{1+}^2+\lambda_{1-}^2\right)x_2^{(2)}
	+\lambda_{1+}^2\lambda_{1-}^2x_2&=0,
\end{align}
where $x_1$ and $x_2$ are determined from \eqref{eq:cancoord}. It is noteworthy to mention that, although the diagonalisation procedure leads to an effective system of two free PUOs, the original system is formed of two interacting PUOs. Moreover, every higher-derivative theory is intrinsically interacting. Now we notice that the system of differential equations \eqref{eq:PUOeqs} is exactly equivalent to the systems \eqref{eq:EoMPUO} and \eqref{eq:EoMPUO2} and we can map both systems. Therefore the occurrence of critical point of the $B$-field can be translated into critical values of the parameters of the system of PUOs. Namely, if
\begin{equation}
	2c_2^2\leq\left[c_0-c_1\pm(\omega_0-\omega_1)\right]^2,
\end{equation}
the frequencies become imaginary, which means instabilities. The above inequality gives a certain range of values for the parameter $c_2$, which is proportional to the strength of the coupling between the PUOs. Similar behaviour is observed in the pp-wave limit of the PW geometry for certain finite set of oscillator modes, where the unstable modes appear for large enough values of the $B$-field \cite{Brecher:2002ar}. In other words, the string can only see the $B$-field if it is excited. In studying supergravity backgrounds one normally requires small curvature of the spacetime at string scale. If the curvature is no longer negligible, then the string dynamics goes beyond the supergravity approximation, which forces one to consider the full string theory.

Consequently, holography teaches us that a change in the background/gravity (due to instabilities) changes the dual gauge theory. From field theory point of view such changes could be interpreted as a phase transition.

In the end of the day we would like to look ahead by slightly touching the issue of information geometry. The entanglement entropy naturally generates an emergent parameter space equipped with Riemannian metric. This metric represents the celebrated Fisher information metric, which can be expressed as second derivative of the entanglement entropy \cite{Matsueda:2014lda,Matsueda:2014yza}, $g_{\mu\nu}(K_1,K_2)=\partial_\mu\partial_\nu S(K_1,K_2)$, where $\partial_\mu=\frac{\partial}{\partial K_\mu}$ with $\mu=1,2$. The absence of minus sign in the definition of the Fisher metric accounts for the Riemannian nature of the metric, the reason being the specific type of variables we have used, i.e, the inverse scaled temperatures $K_1$ and $K_2$. For the case under consideration in this paper, the Fisher information metric for the fourth-order PUO takes the form:
\begin{align}
	g_{11}&=\frac{1}{64}k_\text{B}\coth\frac{K_2}{4}\csch^2\frac{K_1}{4}\left[K_1\left(3+5\coth^2\frac{K_1}{4}
	+7\csch^2\frac{K_1}{4}\right)+4\tanh\frac{K_1}{4}\right.\nonumber\\
	&\phantom{{}={}}\left.+4\coth\frac{K_1}{4}\left(K_1+K_2-5+K_2\coth\frac{K_2}{4}
	-2\log\left[\left(e^{K_1}-1\right)\left(e^{K_2}-1\right)\right]\right)\right],\\
	g_{12}&=\frac{1}{32}k_\text{B}\csch^2\frac{K_1}{4}\csch^2\frac{K_2}{4}
	\left[K_1\left(1+2\coth\frac{K_1}{4}\right)+K_2\left(1+2\coth\frac{K_2}{4}\right)-4\right.\nonumber\\
	&\phantom{{}={}}\left.-2\log\left[\left(e^{K_1}-1\right)\left(e^{K_2}-1\right)\right]\vphantom{\coth\frac{K_2}{4}}\right],
\end{align}
and the other two metric components are obtained by symmetry.

Previously we related the instabilities caused by the presence of critical point of the $B$-field to instabilities in the system of PUOs, and argued that the latter may be interpreted as phase transitions. As a consequence of the instabilities, $K_1$ and $K_2$ in the Fisher metric become imaginary, which transforms the hyperbolic functions into trigonometric ones. Roughly speaking, the hyperbolas ``transform into'' circles, which brings the notion of topological order. From holography point of view, the Pilch-Warner solution is one of the most complicated supergravity solutions, thus being very hard instabilities to be investigated directly using entanglement entropy and alike tools. The holography on the other hand, and the AdS/CFT correspondence as a particular example, is a very powerful tool allowing such instabilities to be examined in the field theory, where the entanglement entropy is much better understood and more easily treatable.

In conclusion, the framework of thermo-field dynamics has proved to be very useful for studying the quantum entanglement of higher-derivative theories, the Pais-Uhlenbeck oscillator being an example. As a possible continuation of this work, it would be interesting to use the quantum PUO for analysing different models in string theory by making use of holography. Such microscopic description could contribute significantly to our understanding of the nature of different high-energy phenomena, which are of great interest to modern theoretical and experimental physics.

\section*{Acknowledgements}

We are grateful to Ivan Masterov for valuable comments on the manuscript. This work was supported by the Bulgarian NSF grant T02/6. SM and RR were also supported by Sofia University Research Fund grant \textnumero~85/2016.


\appendix

\section{Block circulant matrices}
\label{sec:BCM}

A block circulant matrix $\mathbf{B}\in\mathcal{BC}_{n,\kappa}$ is defined as
\begin{equation}
	\mathbf{B}=\text{bcirc}(\mathbf{b}_0,\mathbf{b}_1,\ldots,\mathbf{b}_{n-1})\stackrel{\text{def}}{=}
	\begin{pmatrix}
		\mathbf{b}_0 & \mathbf{b}_1 & \cdots & \mathbf{b}_{n-1}\\
		\mathbf{b}_{n-1} & \mathbf{b}_0 & \cdots & \mathbf{b}_{n-2}\\
		\vdots & \vdots & \ddots & \vdots\\
		\mathbf{b}_1 & \mathbf{b}_2 & \cdots & \mathbf{b}_0
	\end{pmatrix},
\end{equation}
where every block $\mathbf{b}_j$, $j=0,1,\ldots,n-1$, is a square $\kappa\times\kappa$ matrix and $\mathcal{BC}_{n,\kappa}$ is the set of all block circulant matrices of dimension $n\kappa\times n\kappa$. The compound vector
\begin{equation}
\label{eq:eigenvecw}
	\mathbf{w}=
	\begin{pmatrix}
		\mathbf{v}\\
		\rho\mathbf{v}\\
		\vdots\\
		\rho^{n-1}\mathbf{v}
	\end{pmatrix},
\end{equation}
where $\mathbf{v}$ is arbitrary non-null $\kappa$-vector and $\rho$ is any of the $n$ roots of 1, $\rho_j=e^{i2\pi j/n}$ solves the eigenvector problem with eigenvalue $\lambda$:
\begin{equation}
\label{eq:eigensysw}
	\mathbf{B}\mathbf{w}=\lambda\mathbf{w}.
\end{equation}
One can readily show that the system \eqref{eq:eigensysw} is equivalent to the system
\begin{equation}
	\mathbf{H}\mathbf{v}=\lambda\mathbf{v},
\end{equation}
where the $\kappa\times\kappa$ matrix $\mathbf{H}$ has the form
\begin{equation}
	\mathbf{H}=\mathbf{b}_0+\rho\mathbf{b}_1+\cdots+\rho^{n-1}\mathbf{b}_{n-1}.
\end{equation}
In the case of real symmetric block circulant matrix $B$ with all submatrices $\mathbf{b}_j$ themselves symmetric, the matrix $\mathbf{H}_j$, corresponding to the $j$th root of 1 $\rho_j$, takes particularly simple form \cite{Tee:2007}
\begin{equation}
	\mathbf{H}_j=\mathbf{b}_0+2\sum_{f=1}^{h-1}\mathbf{b}_f\cos fj\theta+
	\begin{cases}
		0 & \text{if } n=2h-1\\
		\mathbf{b}_h(-1)^j & \text{if } n=2h
	\end{cases},
\end{equation}
where $\theta=\frac{2\pi}{n}$. Thus defined, each $\mathbf{H}_j$ is a real symmetric matrix, hence it possesses $\kappa$ real orthogonal eigenvectors with corresponding to them real eigenvalues. From the identities $\cos f(n-j)\theta=\cos j\theta$ and (for even $n=2h$) $(-1)^{n-j}=(-1)^j$, it follows that for all $n$ and $1\leq j\leq(n-1)\div2$, $\mathbf{H}_{n-j}=\mathbf{H}_j$. Therefore $\mathbf{H}_{n-j}$ and $\mathbf{H}_j$ have the same system of real eigenvectors and eigenvalues. If those are say $\mathbf{v}$ and $\lambda$, then the matrix $\mathbf{B}$ has a pair two complex conjugate eigenvectors \eqref{eq:eigenvecw} corresponding to the double eigenvalue $\lambda$:
\begin{equation}
	\mathbf{w}=
	\begin{pmatrix}
		\mathbf{v}\\
		\rho_j\mathbf{v}\\
		\vdots\\
		\rho_j^{n-1}\mathbf{v}
	\end{pmatrix},
	\quad
	\overline{\mathbf{w}}=
	\begin{pmatrix}
		\mathbf{v}\\
		\overline{\rho_j}\mathbf{v}\\
		\vdots\\
		\overline{\rho_j^{n-1}}\mathbf{v}
	\end{pmatrix}.
\end{equation}
Since the matrix $\mathbf{B}$ is real and symmetric itself, these two eigenvectors must be replaced by their linear combinations $(\mathbf{w}+\overline{\mathbf{w}})/2$ and $i(\mathbf{w}-\overline{\mathbf{w}})/2$:
\begin{equation}
	\mathcal{R}(\mathbf{w})=
	\begin{pmatrix}
		\mathbf{v}\\
		\cos j\theta\,\mathbf{v}\\
		\vdots\\
		\cos(n-1)j\theta\,\mathbf{v}
	\end{pmatrix},
	\quad
	\mathcal{I}(\mathbf{w})=
	\begin{pmatrix}
		\mathbf{v}\\
		\sin j\theta\,\mathbf{v}\\
		\vdots\\
		\sin(n-1)j\theta\,\mathbf{v}
	\end{pmatrix},
\end{equation}
which are orthogonal as well.


\bibliographystyle{utphys}
\bibliography{puo-refs}

\end{document}